# Quota management in dCache or making a perfectly normal file system normal

*Dmitry* Litvintsev[1,*], *Chitrapu* Krishnaveni[2], *Svenja* Meyer[3], *Paul* Millar[3], *Tigran* Mkrtchyan[1], *Lea* Morschel[3], *Albert* Rossi[1], and *Marina* Sahakyan[3]

[1]Fermi National Accelerator Laboratory, PO Box 500. Batavia IL 60510-5011, USA
[2]Linköping University, 581 83 Linköping, Sweden
[3]Deutsches Elektronen-Synchrotron DESY, Notkestraße 85, D-22607 Hamburg, Germany.

**Abstract.** dCache (https://dcache.org) is a highly scalable storage system providing location-independent access to data. The data are stored across multiple data servers as complete files presented to the end-user via a single-rooted namespace. From its inception, dCache has been designed as a caching disk buffer to a tertiary tape storage system with the assumption that the latter has virtually unlimited capacity. dCache can also be configured as a disk-only storage system with no tape backend. Owing to the idea that a tape resource is infinite, or purely physically limited by budget considerations, the system has never provided for any restrictions on how much data can be stored on tape. Likewise, in the disk-only configuration, the capacity of the system is only limited by the aggregate disk capacity of the data servers. In a multi-user environment, however, this has become problematic. This presentation will describe the design and implementation of a user- and group-based quota system, that allows to manage tape and disk space allocations, as part of dCache namespace.

## 1 Introduction

From its inception dCache was designed as a caching disk buffer to a tertiary tape storage system with an assumption that the latter has virtually unlimited capacity [1]. No provisions for storage quota were made. In disk-only (or hybrid) configuration the need to manage space usage by users and groups has come to the fore. Here we present details of implementation of user and group quotas in dCache.

## 2 dCache in a nutshell

dCache is architected as an ensemble of micro-services implemented in Java and communicating with each other by messages over TCP/IP network. See figure 1. Each micro-service performs specific functions and together they act in concert to deliver data to/from clients over the network. The core components of the system are:

- I/O portals, or doors that implement a specific protocol - (G)FTP, WebDAV, NFS, XRootD etc;

---

[*]e-mail: litvinse@fnal.gov

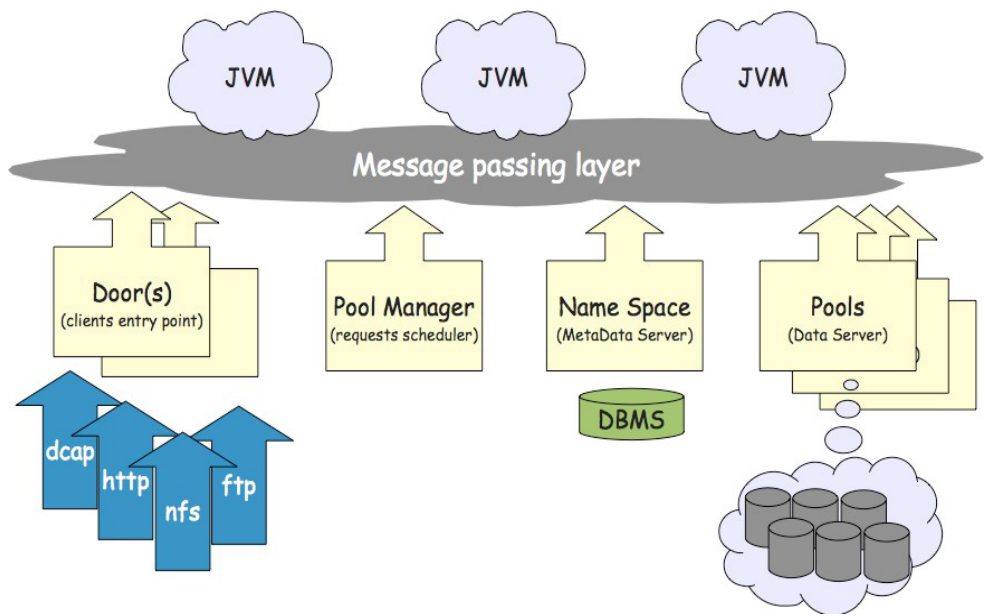

**Figure 1.** dCache in a nutshell. Multiple dCache services run in JVM distributed across the network. The services implement RPC exchanging instructions via messaging layer.

- namespace (also known as PnfsManager) that uses database back-end to store file metadata, directory structure and file locations;
- PoolManager that selects destination (or source) data server based on various selection criteria and pool cost function;
- pool(s) – data servers that store data in complete files (called file replicas) on data partitions, deliver data to/from clients and manage file repository cache. File replicas on pools play the role of inodes to the dCache namespace.

On upload a client connects to a door to initiate a transfer to a destination path in dCache namespace. The door sends a message to the namespace server to check permissions in the destination directory and receives a reply. If OK, a namespace entry is created and the door messages the PoolManager to select a pool. The PoolManager selects the pool based on the pool cost matrix and sends the message to the selected pool to start a mover. The pool starts the mover, replies to the PoolManager with the mover socket address. The PoolManager replies to the door with a message containing the mover address. The door replies to the client with the mover address and the client writes data to that socket.

Following the WLCG SRM specification the lifecycle of file replicas in dCache is determined by their Access Latency/Retention Policy. ONLINE/REPLICA files have disk-only replicas that are always in cache. NEARLINE/CUSTODIAL files have at least one replica stored on tape. Once stored on tape, the cached replicas are subject to cache expungement based on LRU access time.

## 3 Requirements

All objects in the dCache namespace are owned by users identified by UIDs and GIDs. The UID and GID of a dCache user is determined by the mapping of credentials presented by the client to the dCache authn/authz service. The requirements to storage quota system reflect budgetary and fair use concerns. Experiments have requested:

· to be able to control the amount of data written to tape by a user/group;

· to be able to control the amount of disk-only space that a user/group can use;

· quota check and aggregation must not slow down per file namespace operations.

The dCache quota system discussed in this paper has been implemented to address these concerns.

## 4 Implementation

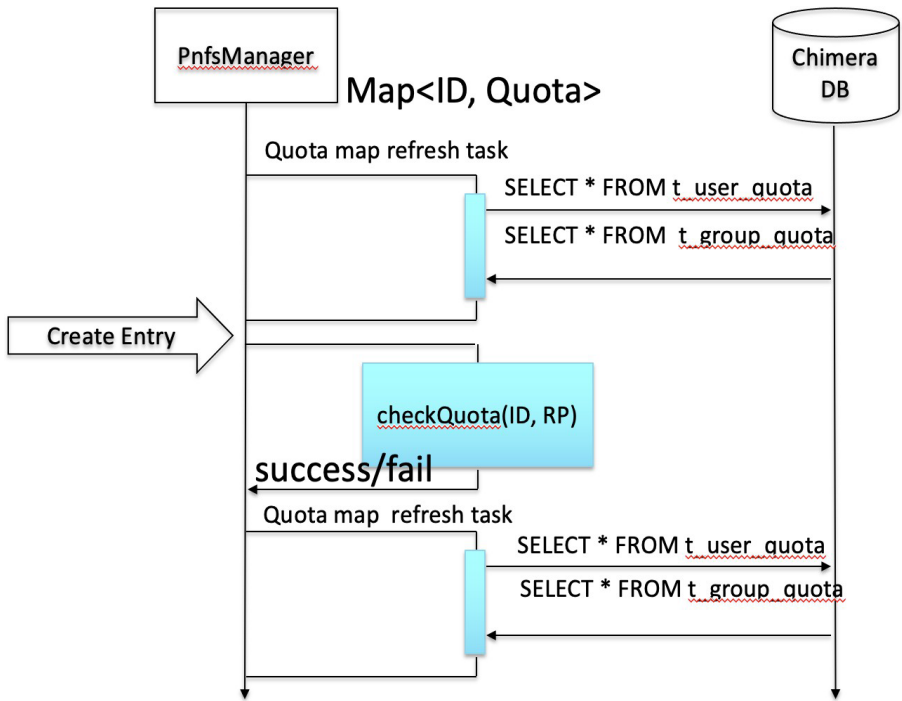

**Figure 2.** Quota handler call sequence.

A quota handler has been added to PnfsManager and it uses the same database back-end for persistency. It implements put/get/modify/remove functions that allow to manipulate user and group quota limits on a data store which is part of the namespace database schema. It functions as follows:

· Schedules periodic scans of all namespace entries to aggregate file sizes by UID, GID and Retention Policy.

  – Updates underlying data store with the resulting usage data.
  – Maintains memory a cache of aggregated data as map<id, Quota>. The Quota data structure contains space usage numbers and their limits for a given Retention Policy (REPLICA, CUSTODIAL, OUTPUT).

- When PnfsManager executes the create entry function, the quota handler checks if used space for a given UID, GID and Retention Policy does not exceed a limit. If it does, the create entry call fails with a "Quota exceeded" message.

The quota system calling sequence is presented in figure 2. The periodic collection of usage data has been chosen to avoid an adverse impact on the performance of per-file operations that would be especially noticeable when performing bulk operations interactively (e.g. rm *) if the usage space were to be re-calculated on each create/remove. The downside of this solution is that the quota usage numbers are always lagging behind. The severity of the lag can be mitigated by adjusting the frequency of the quota aggregation scans. If the user goes over quota the effect is not immediate and that user will still be able to write until the next scan completes and the usage numbers update. Conversely, once over quota, removal of the data will not enable the user to resume writing until the next scan has completed.

## 5 Access to the dCache quota system

### 5.1 Administrative interface

The PnfsManager administrative interface has been augmented with remove/show/set commands allowing the admin to manipulate user and group quota limits:

remove group quota <gid>  # remove group quota
remove user quota <uid>  # remove user quota
set group quota [OPTIONS] <gid> # Set group quota
set user quota [OPTIONS] <uid> # Set user quota
show group quota [-gid=<string>] [-h]  # Print group quota
show user quota [-h] [-uid=<string>]   # Print user quota

### 5.2 Quota REST API

**Figure 3.** dCache quota REST API visualized with the help of the Swagger documentation tool.

The quota REST API has been incorporated into the dCache REST API allowing authenticated users to query their and others' user/group quotas by calling the GET methods of the

API. In addition, the authorized users are allowed to execute POST/PATCH/DELETE to add, modify or remove quotas for a given UID or GID.

The API, documented on the standard dCache REST API port (https://example.org:3880/api/v1) using the Swagger UI [2], is shown in figure 3.

### 5.3 dCache View

When logged into the dCache portal, the user is presented with a web-based GUI that allows interaction with the system. This GUI has been extended to display user UID- and GID-based quotas as pie charts. A color-coded alarm makes it easy to recognize if user is over quota.

## 6 Conclusion

We have implemented storage quotas in dCache which allow control over how much disk-only and tape-resident data a user/group can store in the system. Multi-user, multi-VO installations are expected to benefit from this feature. The quota feature has been rolled out in version 7.2 of dCache where it was not enabled by default. In version 8.2 it is enabled by default.

## 7 Acknowledgments

This work is supported by the US-CMS HL-LHC R&D initiative, and Fermi Research Alliance, LLC under Contract No. DE-AC02-07CH11359 with the U.S. Department of Energy, Office of Science, Office of High Energy Physics.